\documentclass[prd,showpacs,showkeys,nofootinbib,twocolumn]{revtex4-1}

\usepackage{amsmath,amsfonts,amssymb}

\begin{document}

\def\ut{{\underset {\widetilde{\ \ }}u}}
\def\at{{\underset {\widetilde{\ \ }}a}}
\def\ap{\check{a}}

\title{Constitutive law of nonlocal gravity}

\author{Dirk Puetzfeld}
\email{dirk.puetzfeld@zarm.uni-bremen.de}
\homepage{http://puetzfeld.org}
\affiliation{ZARM, University of Bremen, Am Fallturm, 28359 Bremen, Germany} 

\author{Yuri N. Obukhov}
\email{obukhov@ibrae.ac.ru}
\affiliation{Theoretical Physics Laboratory, Nuclear Safety Institute, 
Russian Academy of Sciences, B.Tulskaya 52, 115191 Moscow, Russia} 

\author{Friedrich W. Hehl}
\email{hehl@thp.uni-koeln.de}
\affiliation{Institute for Theoretical Physics, University of Cologne, 50923 Cologne, Germany}

\date{ \today}

\begin{abstract}
We analyze the structure of a recent nonlocal generalization of Einstein's theory of gravitation by Mashhoon et al. By means of a covariant technique, we derive an expanded version of the nonlocality tensor which constitutes the theory. At the lowest orders of approximation, this leads to a simplification which sheds light on the fundamental structure of the theory and may prove useful in the search for exact solutions of nonlocal gravity.
\end{abstract}


\maketitle


\section{Introduction}\label{introduction_sec}

In a series of works \cite{Mashhoon:1993:1,Chicone:Mashhoon:2002:1,Chicone:Mashhoon:2002:2,Mashhoon:2003:1,Mashhoon:2004:1,Mashhoon:2005:1,Mashhoon:2005:2,Mashhoon:2007:1,Mashhoon:2007:2,Mashhoon:2007:3,Chicone:Mashhoon:2007:1,Mashhoon:2008:1,Hehl:Mashhoon:2009:1,Hehl:Mashhoon:2009:2,Mashhoon:2009:1,Blome:etal:2010:1,Mashhoon:2011:1,Mashhoon:2011:2,Chicone:Mashhoon:2012:1,Mashhoon:2013:1,Mashhoon:2013:2,Rahvar:Mashhoon:2014:1,Mashhoon:2014:1,Mashhoon:2015:1,Chicone:Mashhoon:2016:1,Chicone:Mashhoon:2016:2,Mashhoon:2016:1,Bini:Mashhoon:2016:1,Roshan:Rahvar:2019}, eventually culminating in the book \cite{Mashhoon:2017}, Mashhoon and collaborators proposed a nonlocal extension to Einstein's gravity termed nonlocal gravity (NLcG). 

In NLcG gravity is assumed to be history dependent, i.e.\ the gravitational interaction has an additional feature of nonlocality in the sense of an influence (``memory'') from the past that endures. The theory is built upon an ansatz for the so-called nonlocality tensor $N_{ijk}$, leading to a set of integro-differential field equations. The complexity of these equations surpasses the complexity of the ones encountered in Einstein's theory of gravity by a great deal. This makes the search for exact or even approximate solutions of NLcG a daunting task, even if additional symmetry assumptions are made. It is this complexity at a fundamental level, which makes nonlocal gravity a theory for which no exact solutions are known beyond the flat case, in other words, no exact solution encompassing a gravitational field is known. At the same time much work was put into the linearized version of the theory \cite{Mashhoon:2014:1}, and in this context some very promising properties of NLcG -- for example addressing the dark matter problem \cite{Hehl:Mashhoon:2009:1,Blome:etal:2010:1,Chicone:Mashhoon:2012:1,Rahvar:Mashhoon:2014:1,Roshan:Rahvar:2019} -- have been worked out.

However, the fact that no exact solutions exist should of course be remedied, for the theory is supposed to be the successor of General Relativity (GR), for which several solutions are known, which in turn play a key role in the conceptual understanding of the theory. In the present work we show, that the initial choice for the nonlocality is not the ``simplest expression'' for $N_{ijk}$, contrary to what is stated in \cite[(6.107)]{Mashhoon:2017}. We hope that our simplification will pave the way towards a more manageable version of the theory, yet retaining its compelling overall structure.

The structure of the paper is as follows: In section \ref{sec_nonlocal_gravity} we summarize the main features of NLcG. This is followed by a brief review of a covariant expansion technique in \ref{sec_covariant_expansions}. This technique is then applied in \ref{sec_simplified_kernel} to derive an expanded and thereby simplified version of the nonlocality tensor. We conclude our paper in section \ref{sec_conclusions} with a discussion and outlook. An overview of our notation can be found in table \ref{tab_symbols} in appendix \ref{sec_notation}. 

\section{Nonlocal gravity as a teleparallel gravity theory}\label{sec_nonlocal_gravity}

Originally Mashhoon tried to implement the generalization of the locality principle directly for the field equations of GR. This did not appear to be feasible, and a successful starting point turned out to be a rather particular translational gauge theory of gravity (TG), namely the so-called teleparallel equivalent GR$_{||}$ of Einstein's GR, see \cite{Hehl:etal:1980,Hehl:1980,Aldrovandi:Pereira:2013}. 

For a TG, the translational gauge field potential is represented by the coframe $e_i{}^\alpha$, the translational gauge field strength by its covariant ``curl'', the torsion of spacetime:
\begin{equation}
T_{ij}{}^\alpha:=2\left(\partial_{[i}e_{j]}{}^\alpha+\Gamma_{[i|\beta}{}^\alpha e_{|j]}{}^\beta\right).\label{Omega}
\end{equation}
Here coordinate indices are denoted by $i,j,k,...=0,1,2,3$, and frame indices by $\alpha,\beta,\gamma,...=0,1,2,3$, and $\Gamma_{i\alpha}{}^\beta$ is the Lorentz connection of spacetime, see \cite{Blagojevich:Hehl:2013}. The curvature tensor of the spacetime vanishes,
\begin{equation}
R_{ij\alpha}{}^\beta(\Gamma) = 2 \partial_{[i} \Gamma_{j]\alpha}{}^\beta
+ 2 \Gamma_{[i|\gamma}{}^\beta\Gamma_{j]\alpha}{}^\gamma = 0,\label{Rzero}
\end{equation}
that is, we have a teleparallelism, and we can pick a global gauge -- which is denoted by a star over the equality sign -- such that at each point in spacetime the Lorentz connection $\Gamma_{i}{}^{\alpha\beta}=-\Gamma_{i}{}^{\beta\alpha}$ vanishes.

The inhomogeneous and the homogeneous gravitational field equations of NLcG have a Maxwellian structure, see \cite[(5.70) and (6.117)]{Mashhoon:2017}, and are given by 
\begin{eqnarray}
  \partial_j\check{\cal H}^{ij}{}_\alpha-{\cal E}_\alpha{}^i &\stackrel{*}{=}& {\cal T}_\alpha{}^i, \label{fieldeq_1} \\
	\partial_{[i} T_{jk]}{}^\alpha &\stackrel{*}{=}& 0. \label{fieldeq_2}
\end{eqnarray}
The gravitational excitation $\check{\cal H}^{ij}{}_\alpha$, in a Lagrange-Hamilton picture, is the ``momentum'' conjugate to the ``coordinate'' $e_i{}^\alpha$, and the ``velocity'' $ T_{ij}{}^\alpha$: $\check{\cal H}^{ij}{}_\alpha := -2{\partial {\cal L}_{\rm g}} / {\partial T_{ij}{}^\alpha}$, where ${\cal L}_{\rm g}$ is the gravitational Lagrangian density.

The nonlinear correction terms in (\ref{fieldeq_1}) represent the energy-momentum tensor density of the gravitational gauge field,
\begin{equation}
{\cal E}_\alpha{}^i:=-\frac 14  e^i{}_\alpha( T_{ jk}{}^\beta \check{\cal H}^{jk}{}_\beta) + T_{\alpha k}{}^\beta \check{\cal H}^{ik}{}_\beta.\label{energy}
\end{equation}
As source, we have on the right-hand side of the inhomogeneous field equation (\ref{fieldeq_1}) the energy-momentum tensor density of matter ${\cal T}_{\alpha}{}^i$. It has to be assumed symmetric, ${\cal T}_{[\alpha\beta]}=0$, cf.\ \cite[page 52, 1st paragraph, eqs.\ (4.42), (4.43) and (4.36)]{Hehl:1980}.

In a local and linear TG one assumes, as usual in a gauge theory, that the gravitational Lagrangian is quadratic in the field strength -- here in the form of the torsion. Thus, the constitutive law between excitation and field strength is local and linear:
\begin{equation}
\check{\cal H}^{ij}{}_{\alpha} = \frac{1}{2}\chi^{ij}{}_\alpha{}^{kl}{}_\beta\, T_{kl}{}^\beta.\label{loclin}
\end{equation}
General relativity is recovered, see \cite{Cho:1976:1}, via 
\begin{eqnarray}
{}^{\text{GR$_{||}$}}\chi^{ij}{}_{m}{}^{kl}{}_{n}(g)&=&\frac{\sqrt{-g}}{\varkappa}\left(-\,g^{k[i}g^{j]l}g_{mn}\right.\nonumber\\
&&  \hspace{-5pt}\left. -\,4\,\delta^{[i}_{\,\,m}g^{b][k}\delta^{l]}_{n}+ 2\,\delta^{[i}_{\,\,n} g^{j][k}\delta_{m}^{l]}\right),\label{even_endresult}
\end{eqnarray}
where $g^{ij}$ denotes the metric of spacetime, with signature $(+1,-1,-1,-1)$, and $\varkappa$ is Einstein's gravitational constant. 

A nonlocal generalization of a gravity theory is determined by an ansatz, which no longer necessarily can be derived from a Lagrangian, 
\begin{eqnarray}
&& \check{\cal H}^{y_1 y_2}{}_{\upsilon_3} = \frac{1}{2} \Bigg[\chi^{y_1 y_2}{}_{\upsilon_3}{}^{y_4 y_5}{}_{\upsilon_5} \, T_{y_4 y_5}{}^{\upsilon_5} \nonumber \nonumber \\
&&  -\int \sigma^{y_1}{}_{x_1}\sigma^{y_2}{}_{x_2} \sigma_{\upsilon_3}{}^{\xi_3} {\mathcal K}(x,y) X^{x_1 x_2}{}_{\xi_3}{}^{x_4 x_5}{}_{\xi_6}\,T_{x_4 x_5}{}^{\xi_6}d^4x\Bigg],\nonumber \\ \label{excitation}
\end{eqnarray}
where the integration is performed over a 4-dimensional volume; see \cite[(6.114)]{Mashhoon:2017}. Here we use a condensed notation (common to the theory of bitensors) in which the point to which the index of a bitensor belongs can be directly read from the index itself; e.g., $y_n$ denotes indices at the spacetime point $y$. Moreover, in order to distinguish the local frame indices, we use $\xi_1, \xi_2, \dots$ and $\upsilon_1, \upsilon_2, \dots$ to designate objects with frame indices at the point $x$ or $y$, in complete analogy to the labels $x_1, x_2, \dots$ and $y_1, y_2, \dots$ used in the holonomic case. 

In the early works \cite{Hehl:Mashhoon:2009:1,Hehl:Mashhoon:2009:2} it was suggested to use 
\begin{equation}
  \chi^{ij}{}_\alpha{}^{kl}{}_\beta\equiv X^{ij}{}_{\alpha}{}^{kl}{}_{\beta} \equiv {}^{\text{GR$_{||}$}}\chi^{ij}{}_{\alpha}{}^{kl}{}_{\beta} \label{NLG2009}
\end{equation}
as an ansatz for the nonlocal theory. However, it was later on \cite{Mashhoon:2014:1} generalized to 
\begin{eqnarray}
 \chi^{ij}{}_\alpha{}^{kl}{}_\beta&\equiv&{}^{\text{GR$_{||}$}}\chi^{ij}{}_{\alpha}{}^{kl}{}_{\beta},\label{NLG2014_0}\\
X^{ij}{}_\alpha{}^{kl}{}_\beta&\equiv{}&^{\text{GR$_{||}$}}\chi^{ij}{}_{\alpha}{}^{kl}{}_{\beta}+{}^{\text{odd}}\chi^{ij}{}_{\alpha}{}^{kl}{}_{\beta},\label{NLG2014}
\end{eqnarray}
so that
\begin{equation}
{}^{\text{odd}}\chi^{ij}{}_{\alpha}{}^{kl}{}_{\beta}T_{kl}{}^\beta \sim \check{p}\left(
\check{T}{}^ie^j{}_\alpha - \check{T}{}^je^i{}_\alpha\right),\label{XS}
\end{equation}
with a new parity-odd coupling parameter $\check{p}$, see also \cite[(6.109)]{Mashhoon:2017}, that controls the contribution of the axial torsion which is defined as
\begin{equation}
\check{T}{}_i := \frac{1}{3}\,\eta_{ijkl}\,T^{jkl}.\label{cS}
\end{equation}
Another possibility -- which, however, was not followed up -- would be the additional term ${}^{\text{odd}}\chi^{ij}{}_{\alpha}{}^{kl}{}_{\beta}$ on the right-hand side of eq.\ (\ref{NLG2014_0}). 

Recently, a thorough analysis of the most general linear local constitutive relations in the teleparallel gravity has been performed in \cite{Itin:etal:2018}, focusing mainly on its irreducible decomposition. One can prove that a general metric-dependent parity-odd part of the constitutive tensor reads
\begin{eqnarray}
{}^{\text{odd}}\chi^{ij}{}_{\alpha}{}^{kl}{}_{\beta}(g) &&= \frac{\sqrt{-g}}{\varkappa}\left[ \beta_4\,\eta^{ijkl}\,g_{\alpha\beta} + \beta_5\,\eta^{ij[k}{}_{[\alpha}\,e^{l]}{}_{\beta]}\right.\nonumber\\
 &&\left. + \beta_6\left(e^{[i}{}_{(\alpha}\eta^{j]kl}{}_{\beta)}- e^{[k}{}_{(\alpha}\eta^{l]ij}{}_{\beta)}\right)\right].\label{odd_endresult}
\end{eqnarray}     
Accordingly, Mashhoon's constitutive tensor (\ref{NLG2014})-(\ref{XS}) encompasses all 6 irreducible parts (principal, skewon and axion, both even and odd parities), and the corresponding coupling constants of this irreducible decomposition read: $\beta_1 = -1, \beta_2 = -4, \beta_3 = 2, \beta_4 = -\check{p}/6, \beta_5 = -2\check{p}/3, \beta_6 = \check{p}/3$. For the complete notational and computational details see \cite{Itin:etal:2018}; note though that there is a conventional overall factor between our and Mashhoon's coupling constants, and a difference in the definition of the torsion. It is particularly noteworthy that the constitutive relation in general contains a nontrivial skewon part which means that such a constitutive law is not reversible and therefore cannot be derived from a variational principle. For a general introduction to the underlying premetric framework of electrodynamics and gravity see \cite{Hehl:Obukhov:2001:1,Muench:Hehl:Mashhoon:2000:1,Hehl:Obukhov:2003}.

Defining the tensors
\begin{eqnarray}
X^{ij}{}_k &:=& {\frac 12}X^{ij}{}_k{}^{pq}{}_r\,T_{pq}{}^r,\label{XX} \\
\chi^{ij}{}_k &:=& {\frac 12}\chi^{ij}{}_k{}^{pq}{}_r\,T_{pq}{}^r, \label{chiX}
\end{eqnarray}
and by switching to holonomic coordinates, we can recast (\ref{excitation}) into
\begin{equation}
\check{\mathcal H}^{y_1 y_2}{}_{y_3} = \chi^{y_1 y_2}{}_{y_3} + N^{y_1 y_2}{}_{y_3}.\label{excitation_hol}
\end{equation}
Here we introduced 
\begin{eqnarray}
N^{y_1 y_2}{}_{y_3}:=-\int\sigma^{y_1 x_1}\sigma^{y_2 x_2}\sigma_{y_3 x_3} {\mathcal K}(x,y)  X_{x_1 x_2}{}^{x_3} d^4x \nonumber \\
\label{nonlocality_definition}
\end{eqnarray}
for the nonlocal part of (\ref{excitation}), see also \cite[eq.\ (6.107)]{Mashhoon:2017}. In the rest of this work we are going to focus on this nonlocality tensor $N^{y_1 y_2}{}_{y_3}$.

\section{Covariant expansions}\label{sec_covariant_expansions}

In the following we make use of a covariant expansion technique based on a generalization of Synge's ``world function'' $\sigma(x,y)$ \cite{Ruse:1930:1,Synge:1960,DeWitt:Brehme:1960}. Since NLcG is a theory which is based on a non-Riemannian spacetime, we first need to introduce the properties of a world function based on autoparallels in a Riemann-Cartan background. In contrast to a Riemannian spacetime, a Riemann-Cartan spacetime is endowed with an asymmetric connection $\Gamma_{ab}{}^c$, and there will be differences when it comes to the basic properties of a world function $\sigma$ based on autoparallels.

The curvature and the torsion are defined w.r.t.\ the general connection $\Gamma_{ab}{}^c$ as follows:
\begin{eqnarray}
R_{abc}{}^d &:=& 2 \partial_{[a} \Gamma_{b]c}{}^d + 2 \Gamma_{[a|n}{}^d \Gamma_{b]c}{}^n , \label{curvature_def} \\
T_{ab}{}^c &:=& 2 \Gamma_{[ab]}{}^c. \label{torsion_def} 
\end{eqnarray}
The symmetric Levi-Civita connection $\overline{\Gamma}_{kj}{}^i$, as well as all other Riemannian quantities, are denoted by an additional overline. For a general tensor $A$ of rank $(n,l)$ the commutator of the covariant derivative thus takes the form:
\begin{eqnarray}
\left(\nabla_a \nabla_b - \nabla_b \nabla_a \right) A^{c_1 \dots c_n}{}_{d_1 \dots d_l} = - T_{ab}{}^e \nabla_e A^{c_1 \dots c_k}{}_{d_1 \dots d_l} \nonumber \\
+ \sum^k_{i=1} R_{abe}{}^{c_i} A^{c_1 \dots e \dots c_k}{}_{d_1 \dots d_l} - \sum^l_{j=1} R_{abd_j}{}^{e} A^{c_1 \dots c_k}{}_{d_1 \dots e \dots d_l}. \nonumber \\\label{cov_deriv}
\end{eqnarray}
In addition to the torsion, we define the contortion $K_{kj}{}^i$ with the following properties
\begin{eqnarray}
K_{kj}{}^i&:=& \overline{\Gamma}_{kj}{}^i - \Gamma_{kj}{}^i, \label{distorsion_def}\\
K_{kji}&=& - \frac{1}{2} \left( T_{kji} + T_{ikj} + T_{ijk} \right) , \label{torsion_distorsion_1}\\
T_{kj}{}^i&=&-2K_{[kj]}{}^i. \label{torsion_distorsion_2}
\end{eqnarray}

For a world function $\sigma$ based on autoparallels, we have the following basic relations in the case of spacetimes with asymmetric connections:
\begin{eqnarray}
\sigma^x \sigma_x = \sigma^y \sigma_y &=& 2 \sigma, \label{rel_1}\\
\sigma^{x_2} \sigma_{x_2}{}^{x_1} &=& \sigma^{x_1}, \label{rel_2}\\
\sigma_{x_1 x_2} - \sigma_{x_2 x_1} &=& T_{x_1 x_2}{}^{x_3} \partial_{x_3} \sigma.  \label{sigma_commutator}
\end{eqnarray}
We denote higher-order covariant derivatives of the world function by $\sigma^y{}_{x_1\dots y_2\dots}:=\nabla_{x_1}\dots\nabla_{y_2}\dots (\sigma^y)$.

For the covariant expansions we need the limiting behavior of a bitensor $B_{\dots}(x,y)$ when $x$ approaches the reference point $y$. This so-called coincidence limit of a bitensor $B_{\dots}(x,y)$ is a tensor 
\begin{eqnarray}
\left[B_{\dots} \right] = \lim\limits_{x\rightarrow y}\,B_{\dots}(x,y),\label{coin}
\end{eqnarray}
at $y$ and will be denoted by square brackets. In particular, for a bitensor $B$ with arbitrary indices at different points (here just denoted by dots), we have the rule \cite{Synge:1960}
\begin{eqnarray}
\left[B_{\dots} \right]_{;y} = \left[B_{\dots ; y} \right] + \left[B_{\dots ; x} \right]. \label{synges_rule}
\end{eqnarray}
We collect the following useful identities for the world function $\sigma$:
\begin{eqnarray}
{}[\sigma]=[\sigma_x]=[\sigma_y]&=&0 , \label{coinc_1}\\
{}[\sigma_{x_1 x_2}]=[\sigma_{y_1 y_2}]&=&g_{y_1 y_2} , \label{coinc_2}\\
{}[\sigma_{x_1 y_2}]=[\sigma_{y_1 x_2}]&=&-g_{y_1 y_2} , \label{coinc_3}\\
{}[\sigma_{x_3 x_1 x_2}] + [\sigma_{x_2 x_1 x_3}] &=&0. \label{coinc_4}
\end{eqnarray}
Note that up to the second covariant derivative the coincidence limits of the world function match those in spacetimes with symmetric connections. However, at the next (third) order the presence of the torsion leads to
\begin{eqnarray}
{}[\sigma_{x_1 x_2 x_3}] &=& \frac{1}{2} \left(T_{y_1 y_3 y_2} + T_{y_2 y_3 y_1}+ T_{y_1 y_2 y_3} \right) = K_{y_2 y_1 y_3} ,\nonumber\\
\end{eqnarray}
where in the last equality we made use of the contortion $K$. With the help of (\ref{synges_rule}), we can obtain the other combinations with three indices:
\begin{eqnarray}
{}[\sigma_{y_1 x_2 x_3}] = -[\sigma_{y_1 y_2 x_3}] = [\sigma_{y_1 y_2 y_3}] = K_{y_2 y_1 y_3}. \label{coinc_5}
\end{eqnarray}
At the fourth order we have
\begin{eqnarray}
&& K_{y_1}{}^{y}{}_{y_2} K_{y_3 y y_4} + K_{y_1}{}^{y}{}_{y_3} K_{y_2 y y_4} + K_{y_1}{}^{y}{}_{y_4} K_{y_2 y y_4} \nonumber \\
&&+[\sigma_{x_4 x_1 x_2 x_3}] + [\sigma_{x_3 x_1 x_2 x_4}] + [\sigma_{x_2 x_1 x_3 x_4}] =0, \label{coinc_6}\end{eqnarray}
and in particular 
\begin{eqnarray}
[\sigma_{x_1 y_2 y_3 y_4}] &=&   - \frac{1}{3} \nabla_{y_1} \left( K_{y_3 y_4 y_2} + K_{y_2 y_4 y_3}\right)   \nonumber \\
&& + \frac{1}{3} \nabla_{y_3} K_{y_1 y_4 y_2}  + \frac{1}{3} \nabla_{y_2} K_{y_1 y_4 y_3} \nonumber\\
&& +  \nabla_{y_4} K_{y_3 y_1 y_2} - \pi_{y_1 y_4 y_3 y_2},\label{coinc_14}\\ 
\pi_{y_1 y_2 y_3 y_4} &:=& \frac{1}{3} \Big[ K_{y_1 y_2}{}^{y} \left( K_{y_3 y_4 y} + K_{y_4 y_3 y} \right) \nonumber\\
&& -  K_{y_1 y_3}{}^{y} \left( K_{y_4 y_2 y} + K_{y y_2 y_4} \right) \nonumber\\
&&- K_{y_1 y_4}{}^{y} \left( K_{y_3 y_2 y} + K_{y y_2 y_3} \right)  \nonumber \\
&& - 3 K_{y_2 y_1}{}^{y} K_{y_3 y_4 y} + K_{y_3 y_1}{}^{y} K_{y y_2 y_4} \nonumber \\
&& + K_{y_4 y_1}{}^{y} K_{y y_2 y_3} + R_{y_1 y_3 y_2 y_4} \nonumber \\
&&+ R_{y_1 y_4 y_2 y_3} \Big]. \label{pi_def}
\end{eqnarray}
Explicit results for the other index combinations can be found in \cite[eqs.\ (19)-(23)]{Puetzfeld:Obukhov:2018:1}.

Finally, let us collect the basic properties of the so-called parallel propagator $g^{y}{}_{x}:=e^{y}{}_{\alpha} e_{x}{}^{\alpha}$, defined in terms of a parallely propagated tetrad $e^{y}{}_{\alpha}$, which in turn allows for the transport of objects, i.e.\ $V^y=g^y{}_x V^x, \quad  V^{y_1y_2}=g^{y_1}{}_{x_1} g^{y_2}{}_{x_2} V^{x_1x_2}$, etc., along an autoparallel: 
\begin{eqnarray}
g^{y_1}{}_{x} g^{x}{}_{y_2}&=&\delta^{y_1}{}_{y_2}, \quad g^{x_1}{}_{y} g^{y}{}_{x_2}=\delta^{x_1}{}_{x_2}, \label{parallel_1} \\
\sigma^x \nabla_x g^{x_1}{}_{y_1} &=& \sigma^y \nabla_y g^{x_1}{}_{y_1}=0, \nonumber \\
\sigma^x \nabla_x g^{y_1}{}_{x_1} &=& \sigma^y \nabla_y g^{y_1}{}_{x_1}=0, \label{parallel_2} \\
\sigma_x&=&-g^y{}_x \sigma_y, \quad \sigma_y=-g^x{}_y \sigma_x. \label{parallel_3} 
\end{eqnarray}
Note, in particular, the coincidence limits of its derivatives
\begin{eqnarray}
\left[g^{x_0}{}_{y_1} \right] &=& \delta^{y_0}{}_{y_1},  \\
\left[g^{x_0}{}_{y_1 ; x_2} \right] &=& \left[g^{x_0}{}_{y_1 ; y_2} \right] = 0, \label{parallel_4} \\
\left[g^{x_0}{}_{y_1 ; x_2 x_3} \right] &=& - \left[g^{x_0}{}_{y_1 ; x_2 y_3} \right] = \left[g^{x_0}{}_{y_1 ; x_2 x_3} \right] \nonumber \\
&=& - \left[g^{x_0}{}_{y_1 ; y_2 y_3} \right] = \frac{1}{2} R^{y_0}{}_{y_1 y_2 y_3}. \label{parallel_5}
\end{eqnarray}

In the next section we will derive an expanded approximate version of the nonlocality tensor. We make use of the covariant expansion technique \cite{Synge:1960,Poisson:etal:2011} on the basis of the autoparallel world function. For a general bitensor $B_{\dots}$ with a given index structure, we have the following general expansion, up to the third order (in powers of $\sigma^y$):
\begin{eqnarray}
B_{y_1 \dots y_n}&=&A_{y_1 \dots y_n} +  A_{y_1 \dots y_{n+1}} \sigma^{y_{n+1}} \nonumber \\
&& + \frac{1}{2} A_{y_1 \dots y_{n+1} y_{n+2}} \sigma^{y_{n+1}} \sigma^{y_{n+2}} + {\cal O}\left( \sigma^3 \right), \label{expansion_general_yn_1} \\
A_{y_1 \dots y_n}&:=&\left[B_{y_1 \dots y_n}\right] , \label{expansion_general_yn_2} \\
A_{y_1 \dots y_{n+1}}&:=&\left[B_{y_1 \dots y_n ; y_{n+1}}\right] - A_{y_1 \dots y_n ; y_{n+1}} , \label{expansion_general_yn_3} \\
A_{y_1 \dots y_{n+2}}&:=&\left[B_{y_1 \dots y_n ; y_{n+1} y_{n+2}}\right]- A_{y_1 \dots y_n y_0} \left[\sigma^{y_0}{}_{y_{n+1} y_{n+2}}\right] \nonumber \\ 
&&  - A_{y_1 \dots y_n ; y_{n+1} y_{n+2}} - 2 A_{y_1 \dots y_n (y_{n+1} ; y_{n+2})} . \label{expansion_general_yn_4}
\end{eqnarray}
With the help of (\ref{expansion_general_yn_1}) we are able to iteratively expand any bitensor to any order, provided the coincidence limits entering the expansion coefficients can be calculated. We note in passing, that this expansion technique has also been applied extensively in the context of the equations of motion of extended test bodies \cite{Dixon:1964,Dixon:1974,Dixon:1979,Dixon:2008,Puetzfeld:Obukhov:2014:2,Dixon:2015,Obukhov:Puetzfeld:2015:1} and in the gravitational self-force problem \cite{Ottewill:2011,Poisson:etal:2011}.
The expansion for bitensors with mixed index structure can be obtained from transporting the indices in (\ref{expansion_general_yn_1}) by means of the parallel propagator. 

\section{Constitutive law}\label{sec_simplified_kernel}

As was demonstrated in section \ref{sec_nonlocal_gravity}, nonlocal gravity is based on an ansatz for the so-called nonlocality tensor $N^{y_1 y_2 y_3}$, which involves a scalar kernel $\mathcal{K}(x, y)$ and a tensor $X^{x_1 x_2 x_3}$. Albeit the form of $N^{y_1 y_2 y_3}$ given in (\ref{nonlocality_definition}), was declared the ``simplest expression'' for the nonlocality tensor in \cite{Mashhoon:2017}, we observe here that a further simplification can be achieved by performing a covariant expansion of the derivatives of the world function entering (\ref{nonlocality_definition}). 

Utilizing the general expansion technique from (\ref{expansion_general_yn_1})-(\ref{expansion_general_yn_4}) we have for the derivative of the world function around an arbitrary reference world line $Y$.
\begin{eqnarray}
\sigma_{y_1 x_2}&=& -g_{y_1 x_2} + g_{x_2}{}^{y} [\sigma_{y_1 x y_3}] \sigma^{y_3} + \frac{1}{2} \bigg( g_{x_2}{}^{y} [\sigma_{y_1 x y_3 y_4}] \nonumber \\
&& - g_{x_2}{}^{y_2} g_{y_1 y} \left[g_{y_2}{}^{x}{}_{;y_3 y_4}\right] - 2 g_{x_2}{}^{y} [\sigma_{y_1 x (y_3}]_{;y_4)}  \nonumber \\
&& - g_{x_2}{}^{y} [\sigma_{y_1 x y_5}] [\sigma^{y_5}{}_{y_3 y_4}] \bigg) \sigma^{y_3} \sigma^{y_4}  + {\cal O}\left( \sigma^3 \right) . \label{expansion_sigma_yx} 
\end{eqnarray} 
With the results for the coincidence limits worked out in the previous section \ref{sec_covariant_expansions}, we end up with the following explicit expansion of the world function derivative up to the second order:
\begin{eqnarray}
\sigma_{y_1 x_2}&=&-g_{y_1 x_2} + g_{x_2}{}^y K_{y_3 y y_1} \sigma^{y_3} + \frac{1}{2} \sigma^{y_3} \sigma^{y_4} g_{x_2}{}^y\bigg[  \nonumber\\
&&\frac{1}{3} \nabla_{(y_3} K_{|y | y_4) y_1}-\frac{1}{3} \nabla_y K_{(y_3 y_4) y_1}  -  \nabla_{(y_4} K_{y_3) y y_1}\nonumber\\
&&+ K_{y_5 y y_1}  K_{(y_4 y_3)}{}^{y_5}  - \pi_{y(y_4 y_3) y_1}\bigg] +{\cal O}\left( \sigma^3 \right),\label{expansion_explicit_sigma_xy} 
\end{eqnarray}
with 
\begin{eqnarray}
\pi_{y(y_4y_3)y_1} &=& \frac{1}{3} \bigg[ K_{y (y_4}{}^{y'} K_{y_3)y_1 y'} - K_{y (y_3}{}^{y'} K_{|y'| y_4) y_1} \nonumber \\
&&  - K_{y y_1}{}^{y'} K_{(y_3 y_4) y'} - 3 K_{(y_4 | y|}{}^{y'} K_{y_3)y_1 y'} \nonumber \\
&& + K_{(y_3 | y}{}^{y'} K_{y'| y_4) y_1}+ R_{y (y_3 y_4) y_1}\bigg]. \label{pi_symmetrized}
\end{eqnarray}
Inserting (\ref{pi_symmetrized}) into (\ref{expansion_explicit_sigma_xy}) we end up with 
\begin{eqnarray}
\sigma_{y_1 x_2}&=& - \,g_{y_1 x_2} + g_{x_2}{}^{y} K_{y_3 y y_1} \sigma^{y_3}  \nonumber\\
 &-&\frac{1}{6} \sigma^{y_3} \sigma^{y_4} g_{x_2}{}^y \Bigg[ R_{y (y_3 y_4) y_1} + \kappa_{y(y_3 y_4) y_1}\Bigg] +{\cal O}\left( \sigma^3 \right), \nonumber\\ \label{expansion_explicit_sigma_xy_explicit} 
\end{eqnarray}
where we collected all contortion terms in the auxiliary variable 
\begin{eqnarray}
\kappa_{y(y_4y_3)y_1} &:=& K_{y (y_4}{}^{y'} K_{y_3)y_1 y'} - K_{y (y_3}{}^{y'} K_{|y'| y_4) y_1} \nonumber \\
&&  - K_{y y_1}{}^{y'} K_{(y_3 y_4) y'} + K_{(y_3 | y}{}^{y'} K_{y'| y_4) y_1}\nonumber \\
&&  - 3 K_{(y_4 | y|}{}^{y'} K_{y_3)y_1 y'} -  3 K_{(y_4 y_3)}{}^{y'} K_{y'yy_1} \nonumber\\
&& + \nabla_y K_{(y_3y_4)y_1} - \nabla_{(y_3} K_{|y|y_4)y_1} \nonumber\\
&& + 3\nabla_{(y_3} K_{y_4)yy_1}\,. \label{kappa}
\end{eqnarray}

\subsection{Riemann-Cartan spacetime}\label{subsec_simplified_kernel_riem-cartan}

Plugging in the expansion from (\ref{expansion_explicit_sigma_xy_explicit}) into the ansatz for the nonlocality (\ref{nonlocality_definition}) we end up with:
\begin{eqnarray}
&&N_{y_1 y_2 y_3}=\int \bigg\{ g_{y_1 x_1} g_{y_2 x_2} g_{y_3 x_3} -  \sigma^{y'} \Big[g_{y_1 x_1} g_{y_2 x_2} g_{x_3}{}^{y} K_{y' y y_3} \nonumber \\ 
&&\phantom{N_{y_1 y_2 y_3}}+ g_{y_2 x_2} g_{y_3 x_3} g_{x_1}{}^{y} K_{y' y y_1} + g_{y_1 x_1} g_{y_3 x_3} g_{x_2}{}^{y} K_{y' y y_2}  \Big]    \nonumber \\
&&+ \frac{1}{6}\sigma^{y_5} \sigma^{y_6} \bigg[ g_{x_1}{}^{y} g_{y_2 x_2} g_{y_3 x_3}\left( 
R_{y (y_5 y_6) y_1} + \kappa_{y(y_5 y_6) y_1}\right)  \nonumber \\
&&\phantom{\sigma^{y_5} \sigma^{y_6} \bigg[} + g_{x_2}{}^{y} g_{y_1 x_1} g_{y_3 x_3}
\left( R_{y (y_5 y_6) y_2} + \kappa_{y(y_5 y_6) y_2}\right)  \nonumber \\
&& \phantom{\sigma^{y_5} \sigma^{y_6} \bigg[} + g_{x_3}{}^{y} g_{y_1 x_1} g_{y_2 x_2}
\left( R_{y (y_5 y_6) y_3} + \kappa_{y(y_5 y_6) y_3}\right)  \nonumber \\
&& \phantom{\sigma^{y_5} \sigma^{y_6} \bigg[} + 6\,g_{y_1 x_1} g_{x_2}{}^{y'} g_{x_3}{}^{y''} 
K_{(y_5| y' y_2|} K_{y_6) y''y_3} \nonumber \\
&& \phantom{\sigma^{y_5} \sigma^{y_6} \bigg[} + 6\,g_{y_2 x_2} g_{x_1}{}^{y'} g_{x_3}{}^{y''} 
K_{(y_5| y' y_1|} K_{y_6) y''y_3} \nonumber \\
&& \phantom{\sigma^{y_5} \sigma^{y_6} \bigg[} + 6\,g_{y_3 x_3} g_{x_1}{}^{y'} g_{x_2}{}^{y''} 
K_{(y_5| y' y_1|} K_{y_6) y''y_2} \bigg] \nonumber \\
&&  \phantom{\sigma^{y_5} \sigma^{y_6} \bigg[} + {\cal O}\left( \sigma^3 \right) \bigg\}\, \mathcal{K}(x,y)  X^{x_1 x_2 x_3}  d^4x. \label{nl_tensor_expanded_riem-cartan} 
\end{eqnarray}
Different orders in this version of the nonlocality (\ref{nonlocality_definition}) correspond to different orders of the approximation in powers of the world function. The expansion (\ref{nl_tensor_expanded_riem-cartan}) clearly exhibits the complicated geometrical structure of the original ansatz (\ref{nonlocality_definition}). The torsion of spacetime, here in the form of the contortion, already enters the picture at the first order. This in turn leads to very complicated field equations of NLcG.   

\subsection{Riemannian spacetime}\label{subsec_simplified_kernel_riem}

Albeit the latest version of nonlocal gravity described in \cite{Mashhoon:2017} uses a Riemann-Cartan spacetime as the geometrical setting, our general method also allows for a direct specialization of (\ref{nonlocality_definition}) to a Riemannian background, i.e.\
\begin{eqnarray}
&&N_{y_1 y_2 y_3}=\int \bigg[ g_{y_1 x_1} g_{y_2 x_2} g_{y_3 x_3} + \frac{1}{6} \bigg(g_{y_1 x_1} g_{y_2 x_2} g_{x_3}{}^{y}  \nonumber \\ 
&&\overline{R}_{y_3 (y' y'') y} + g_{y_2 x_2} g_{y_3 x_3} g_{x_1}{}^{y} \overline{R}_{y_1 (y' y'') y} + g_{y_1 x_1} g_{y_3 x_3} g_{x_2}{}^{y} \nonumber \\
&&\overline{R}_{y_2 (y' y'') y}  \bigg) \sigma^{y'}  \sigma^{y''} + {\mathcal O}(\sigma^4)\bigg] \mathcal{K}(x,y) X^{x_1 x_2 x_3} d^4x . \label{nl_tensor_expanded} 
\end{eqnarray}
Here $\overline{R}$ denotes the Riemannian curvature tensor built from the Levi-Civita connection $\overline{\Gamma}$. In contrast to the Riemann-Cartan case -- in which the torsion entered at the first order -- the expansion (\ref{nl_tensor_expanded}) shows that the specialization to a Riemannian background leads to a mild simplification, in the sense that the geometric terms (i.e.\ the Riemannian curvature) now enter the nonlocality ansatz only at the second order.

\begin{table}
\caption{\label{tab_symbols}Directory of symbols.}
\begin{ruledtabular}
\begin{tabular}{ll}
Symbol & Explanation\\
\hline
&\\
\hline
\multicolumn{2}{l}{{Geometrical quantities}}\\
\hline
$g_{a b}$ & Metric\\
$e_i{}^\alpha$ & Coframe, tetrad\\
$\delta^a_b$ & Kronecker symbol \\
$x^{a}, y^a$ & Coordinates \\
$\eta_{abcd}$ & Totally antisymm.\ Levi-Civita tensor \\ 
$\Gamma_{i\alpha}{}^\beta$ &  Lorentz connection\\
$\Gamma_{a b}{}^c$ & Riemann-Cartan connection \\
$\overline{\Gamma}_{a b}{}^c$ & Levi-Civita connection \\
$R_{a b c}{}^d$ & Curvature \\
$T_{ab}{}^c$ & Torsion\\
$\check{T}{}_a$ & Axial torsion\\
$K_{ab}{}^c$ & Contortion\\
$\sigma(x,y)$ & World function\\
$g^{y_0}{}_{x_0}$ & Parallel propagator\\
&\\
\hline
\multicolumn{2}{l}{{Miscellaneous}}\\
\hline
${\mathcal L}_{\rm g}$ & Gravitational Lagrangian \\
$\check{\mathcal H}^{ij}{}_\alpha$ & Gravitational excitation\\
$\varkappa$ & Gravitational coupling constant\\
${\mathcal E}_\alpha{}^i$ & Gauge field energy-momentum \\
${\cal T}_{\alpha}{}^i$ & Matter energy-momentum \\
$N_{abc}$ & Nonlocality tensor \\
$\mathcal{K}(x,y)$ & Causal kernel \\
$\chi^{ab}{}_c{}^{de}{}_f$ & Constitutive tensor \\
$A_{y_1 \dots y_n}$ & Expansion coefficient \\
$\check{p}$, $\beta_1$, \dots, $\beta_6$ & Coupling parameters\\
$\pi_{y_1 y_2 y_3 y_4}$, $\kappa_{y_1 y_2 y_3 y_4}$ & Auxiliary quantities \\
&\\
\hline
\multicolumn{2}{l}{{Operators}}\\
\hline
$\partial_i$, ``${\phantom{a}}_{,}$'' & Partial derivative \\
$\nabla_i$, ``${\phantom{a}}_{;}$'' & Covariant derivative \\ 
``$[ \dots ]$''& Coincidence limit\\
``$\overline{\phantom{A}}$''& Riemannian object\\
&\\
\end{tabular}
\end{ruledtabular}
\end{table}

\section{Discussion and conclusions}\label{sec_conclusions}

We have worked out an approximate version of the nonlocality ansatz of NLcG by means of a covariant expansion technique. Our results in the Riemann-Cartan (\ref{nl_tensor_expanded_riem-cartan}), as well as in the Riemannian context (\ref{nl_tensor_expanded}), pave the way for a refined version of the theory postulated in \cite{Mashhoon:2017}. A natural improvement can be achieved by using just the lowest order in the expansion (\ref{nl_tensor_expanded_riem-cartan}) as a new basic ansatz for the nonlocality tensor $N_{y_1 y_2 y_3}$. Namely, we propose that the original ansatz (\ref{nonlocality_definition}) should be replaced by
\begin{eqnarray}
N_{y_1 y_2 y_3} =  \int g_{y_1 x_1} g_{y_2 x_2} g_{y_3 x_3} \mathcal{K}(x,y) X^{x_1 x_2 x_3} d^4x. \label{nonlocality_definition_new} 
\end{eqnarray}
This choice provides an essential development of the NLcG theory since it avoids some of the overwhelming geometrical complexity of the original ansatz. At the same time, it is perfectly consistent with all the previous results of NLcG, in particular, it is important that the new ansatz (\ref{nonlocality_definition_new}) is totally compatible with the linearized solutions which have been found so far in the context of NLcG.  

Furthermore, it is worthwhile to note that the new nonlocal constitutive law (\ref{nonlocality_definition_new}) appears to be much more natural from the viewpoint of relativistic multipolar schemes \cite{Puetzfeld:Obukhov:2014:2,Obukhov:Puetzfeld:2015:1} as compared to the original ansatz (\ref{nonlocality_definition}), since it avoids the emergence of derivatives of the world function, which do not have a straightforward interpretation -- in contrast to the appearance of the parallel propagator in the new ansatz (\ref{nonlocality_definition_new}). 

With an account of these advantageous properties, one can expect that our new constitutive law would eventually lead to an exact solution of NLcG, although even with the simplified ansatz for the nonlocality, the solution of the full NLcG field equations still appears to be a daunting task.

\begin{acknowledgments}
We are grateful to Bahram Mashhoon (Tehran) for helpful remarks in the context of nonlocal gravity. Furthermore we thank Yakov Itin (Jerusalem) and Jens Boos (Alberta) for their comments. This work was supported by the Deutsche Forschungsgemeinschaft (DFG) through the Grant No. PU 461/1-1 (D.P.). The work of Y.N.O. was partially supported by PIER (``Partnership for Innovation, Education and Research'' between DESY and Universit\"at Hamburg) and by the Russian Foundation for Basic Research (Grant No. 18-02-40056-mega).
\end{acknowledgments}

\appendix

\section{Notations and conventions}\label{sec_notation}
Table \ref{tab_symbols} contains a brief overview of the symbols used throughout the work.

\bibliographystyle{unsrtnat}
\bibliography{nonlocalgrav_bibliography}
\end{document}